\documentclass[lineno,fleqn,10pt]{wlscirep}
\usepackage[utf8]{inputenc}
\usepackage[T1]{fontenc}

\usepackage{pdflscape}
\usepackage{stix}
\usepackage[section]{placeins}
\usepackage{lineno}

\usepackage{caption}
\usepackage{xr}
\usepackage{setspace}
\usepackage{hyperref}
\usepackage{xr-hyper}
\usepackage{cleveref}
\usepackage{xspace}
\usepackage{stix}
\usepackage{amssymb}

\usepackage{booktabs,tabularx}

\newcommand{\sectionname}{Section}
\newcommand{\etal}{{et al.}\xspace} 
\newcommand{\ie}{{i.e.},\xspace}
\newcommand{\eg}{{e.g.},\xspace}

\newcommand{\matchingdistribution}{\textit{{Matching distribution}}}
\newcommand{\matchingwomen}{\textit{Matching women}}
\newcommand{\matchingmen}{\textit{Matching men}}
\usepackage{booktabs}
\usepackage{threeparttable}
\usepackage{caption}

\makeatletter
\newcommand*{\addFileDependency}[1]{
	\typeout{(#1)}
	\@addtofilelist{#1}
	\IfFileExists{#1}{}{\typeout{No file #1.}}
}
\makeatother

\newcommand*{\myexternaldocument}[2]{%
	\externaldocument[#2]{#1}%
	\addFileDependency{#1.tex}%
	\addFileDependency{#1.aux}%
}

\myexternaldocument{supplementary_revised_3}{S-}

\title{Systematic comparison of gender inequality in scientific rankings across disciplines}

\author[1,5,8,*]{Ana Maria Jaramillo}

\author[2,3,4,*]{Mariana Macedo}

\author[6]{Melanie Oyarzun}

\author[1,7]{Marcos Oliveira}

\author[5,8,+]{Fariba Karimi}

\author[1,9,+]{Ronaldo Menezes}

\affil[1]{Computer Science, University of Exeter, United Kingdom}

\affil[2]{Computer Science and Engineering, Constructor University, Bremen, Germany}
\affil[3]{Data Science Department, Northeastern University London, United Kingdom}
\affil[4]{Khoury College of Computer Sciences, Northeastern University, United States}

\affil[5]{Network Inequality Group, Complexity Science Hub Vienna, Austria}

\affil[6]{Center for Collective Learning, Corvinus Institute
for Advanced Studies at Corvinus University of Budapest, Hungary}

\affil[7]{Computer Science, Vrije Universiteit Amsterdam, Amsterdam, Netherlands} 

\affil[8]{Computational Social Science Group, Technische Universität Graz, Austria} 

\affil[9]{Computer Science, Federal University of Cear\'a, Fortaleza, Brazil} 

\affil[*]{Corresponding authors emails: ajaramillo@biocomplexlab.org, mmacedo@biocomplexlab.org}
\affil[*+]{these authors contributed equally to this work}


\begin{abstract}
Women's participation in academia has grown over recent decades. Yet, it is unclear how this growth translates into representation at the top of academic rankings (measured by scientific productivity and citations). 
Here, we investigate gender gaps in productivity, citations, and coauthorship networks across 18 fields, using 67.7 million papers published between 1975 and 2020 in the Semantic Scholar Open Research Corpus, with authors' binary gender inferred from names using Genderize and Namsor.
We find that women remain consistently underrepresented in top-ranked positions across all fields, even in disciplines where their overall participation is relatively high.
We observe that rankings are generally becoming increasingly rigid over time, with fewer researchers entering or leaving top-ranked positions from one year to the next in most fields, although we do not test whether this rigidity contributes to the disparities we document.
Across fields, highly productive men receive more citations than the closest available women based on the observed productivity and career stages. 
However, when top-ranked women are compared with their closest male counterparts in terms of career stage and publication profile (accounting for venue prestige and authorship position), gaps narrow in some fields and in a few others disappear or reverse, showing that, among highly productive researchers, differences in research output alone do not fully account for the citation gaps.
Our findings suggest that rising participation in academia may not guarantee representation, highlighting the need for more research on the topic and on the design of inclusive ranking systems.
All our findings are limited to the Semantic Scholar Open Research Corpus, the capabilities of the gender inference libraries, and the methodology used in this work; our citation-gap estimates apply only to researchers who have already reached high productivity.
\end{abstract}

\begin{document}

\flushbottom
\maketitle
\thispagestyle{empty}

\maketitle

\section*{Introduction}\label{sec1}

The rising participation of women in academia has reshaped research agendas, from women's health in medicine to the integration of feminist perspectives in policies and peace agreements~\cite{Clayton2016,Blaxill2016,Garcia2024}. 
Yet, this rise has been uneven across fields~\cite{lariviere2013bibliometrics,Huang2020}, and persistent gender gaps remain in productivity~\cite{jaramillo2021reaching}, career length~\cite{glass2013s,Huang2020,jadidi2018gender,MAVRIPLIS2010}, cumulative advantage~\cite{Kong2022}, research agendas~\cite{trusz2020females,ceci2011understanding,beasley2012they}, and pay~\cite{kim2024persistent}.
Yet it remains unclear how women's increasing participation translates into representation at the top of academic rankings (as measured by scientific productivity and citations) and how this relationship varies across disciplines. 

Participation refers to the proportion of a group within a field (\eg 40\% of a field comprises women), while representation refers to visibility in top-ranked positions~\cite{jadidi2018gender}. A large proportion of women in a field does not necessarily equate to their advancement to senior or top-ranked positions in terms of productivity, visibility, and recognition. 
Additionally, because datasets and methodologies differ across studies, there is a lack of comprehensive analyses comparing the Humanities and Social Sciences with the Physical Sciences in terms of gender dynamics and citation advantages. In this work, we address these gaps by investigating gender differences in how participation and representation relate to visibility and performance in academia across various fields, years, and career stages.


Within the social sciences, particularly gender studies and the sociology of science, the study of gender inequalities in academia has been ongoing since the 1980s~\cite{Cocks1988,Rosenfeld1982}. These studies, based primarily on surveys and interviews, demonstrated how `doing gender' (when gender identity is constructed and reinforced through social interactions) reinforces gender stereotypes and divisions of labour, thereby affecting the reduction of inequalities in science. Academia, like other organisational systems, can reinforce inequalities when operating without adequate checks and balances, particularly in hierarchical structures where few reach top positions while those at the bottom perform the most labour-intensive and less prestigious tasks (positions disproportionately occupied by individuals facing intersecting forms of oppression based on gender, class, and race~\cite{acker2006inequality}). While interventions have been implemented to support the careers of women in academia~\cite{Lavere020380}, several barriers hinder women's advancement up the academic ladder, including implicit and explicit gender discrimination~\cite{doering2023me} and the division of labour and power dynamics~\cite{fox2020gender}. As shown for example, in an empirical study conducted in Dutch universities based on hiring committee reports showing that male support networks potentially isolate women from unwritten rules to reach top positions in the humanities, and how, in the natural sciences, all scientists are compared to the `ideal scientist': a tireless worker with a commitment to travel abroad for long periods, which penalises those with household and caring responsibilities (especially women)~\cite{van2012slaying}. 

Current studies using bibliometric databases have consistently found that the gender gap in scientific productivity is due to women having shorter careers than men~\cite{Huang2020, jadidi2018gender, jaramillo2021reaching}. Particularly in Physics and Computer Science, women tend to have smaller and less diverse coauthorship networks than their men counterparts~\cite{jadidi2018gender,jaramillo2021reaching,she2025gender}. These disparities are believed to arise from barriers related to integration, social acceptance, and peer recognition, all of which are exacerbated in fields with a significant underrepresentation of women~\cite{Astegiano2019}. Despite efforts to improve gender diversity in specific disciplines, systematic investigations are still needed to understand how increased participation correlates with women's underrepresentation in top-ranked positions and their shorter career durations across various fields. 

To succeed in academia and attain the status of a top-ranked researcher, it is crucial to maintain high levels of publications, citations, funding, and established coauthorships~\cite{Huang2020}. These achievements result from a combination of individual and collective efforts. However, these efforts do not appear to yield the same benefits for all researchers, with historically underrepresented groups facing thicker glass ceilings~\cite{patton2004reflections,Liu2010RACE,Gomez2022}, as some researchers may need to exert greater effort than others to achieve similar levels of success in academia (e.g. measured by citations or by reaching higher career stages). The question remains whether rising participation is accompanied by representation at the top, and whether comparable scholarly contributions are reflected in comparable success and visibility across genders.



Here, we build, merge, and analyse one of the largest datasets of scientific publications with gender-inferred authors, including 67,696,061 publications (representing 35\% of the total number of publications in Semantic Scholar) and more than 4 million women and 11 million men identified researchers. At first, by studying aggregate rankings of productivity, citations, and coauthorship, we show that despite incremental participation of women up to 2020, senior women researchers accounted for only 25\% of top-ranked positions across disciplines. We also find that all fields exhibit rigid ranking structures, in which the top 1,000 and 1\% of researchers remain relatively stable over time, with fewer researchers entering or leaving these positions over the study period, 
in line with the literature~\cite{Iniguez2022}. 


We then analyse whether those results remain consistent when moving from aggregate rankings to individual-level comparisons. Within each field and career stage, we compare citation gains of highly productive women and men with similar publication profiles, accounting for cohort, productivity, venue prestige, authorship position, and coauthorships. We observe, using Negative Binomial fixed-effects models, that citation differences generally favour men, although the size of these gaps depends on the sampling of top-ranked researchers and fields. As a first benchmark, we compared the top 5\% most productive researchers within each field and career stage, keeping the observed proportions of women and men in the data (\matchingdistribution). Women receive 13.9\% fewer expected citations on average, with significant gaps favouring men when looking at fields together and in 13 of 18 fields, with the largest in Mathematics (16.7\%). For the remaining five fields, differences favour women in Art and History (6.6\% and 5.7\%, respectively) and are not significant in Environmental Science, Geology, and Physics.

We further refine the comparison by anchoring it separately on each gender group. When each highly productive man is paired with the most similar woman in terms of publication profile, within the same field and career stage (\matchingmen), the pattern intensifies: citation gaps favour men in all fields, with women receiving 39.1\% fewer expected citations than men on average, ranging from 9.5\% fewer in Geology to 48.3\% in Mathematics. These gaps broadly follow women’s participation across fields, with the largest gaps in lower-participation pSTEM fields, yet remaining substantial even in higher-participation fields such as Psychology, where women still receive 18.5\% fewer citations. By contrast, when the comparison is anchored on highly productive women and their closest men counterparts (\matchingwomen), citation gaps narrow substantially, averaging 6.5\% fewer citations for women. Men still receive significantly more expected citations in 12 fields, although in 9 of these fields the gap is smaller than in the men-anchored comparison. In the remaining fields, differences are non-significant in Geology and Sociology and favour women in Art (18.4\%), Computer Science (5.1\%), Psychology (2.8\%), and Physics (2.6\%). 

Taken together, all our models, findings, and analyses show that similar publication trajectories are not always accompanied by comparable citation counts across gender groups, whereas the highest levels of citation-based rankings, which are incrementally rigid, continue to show the historical overrepresentation and cumulative recognition of men. The increase in women's participation has not yet been accompanied by a proportional increase in women’s representation, but it hints at possible improvements when researchers have comparable backgrounds across genders, and this may also be true for other sociodemographic groups. Our findings then highlight persistent disparities in representation in top-ranked positions and in citation gains among highly productive researchers.

\section*{Results}


\subsection*{Growth of women's participation}
\label{sec:women_growth_over_time}

\begin{figure}[!ht]
    \centering
\includegraphics[width=0.9\textwidth]{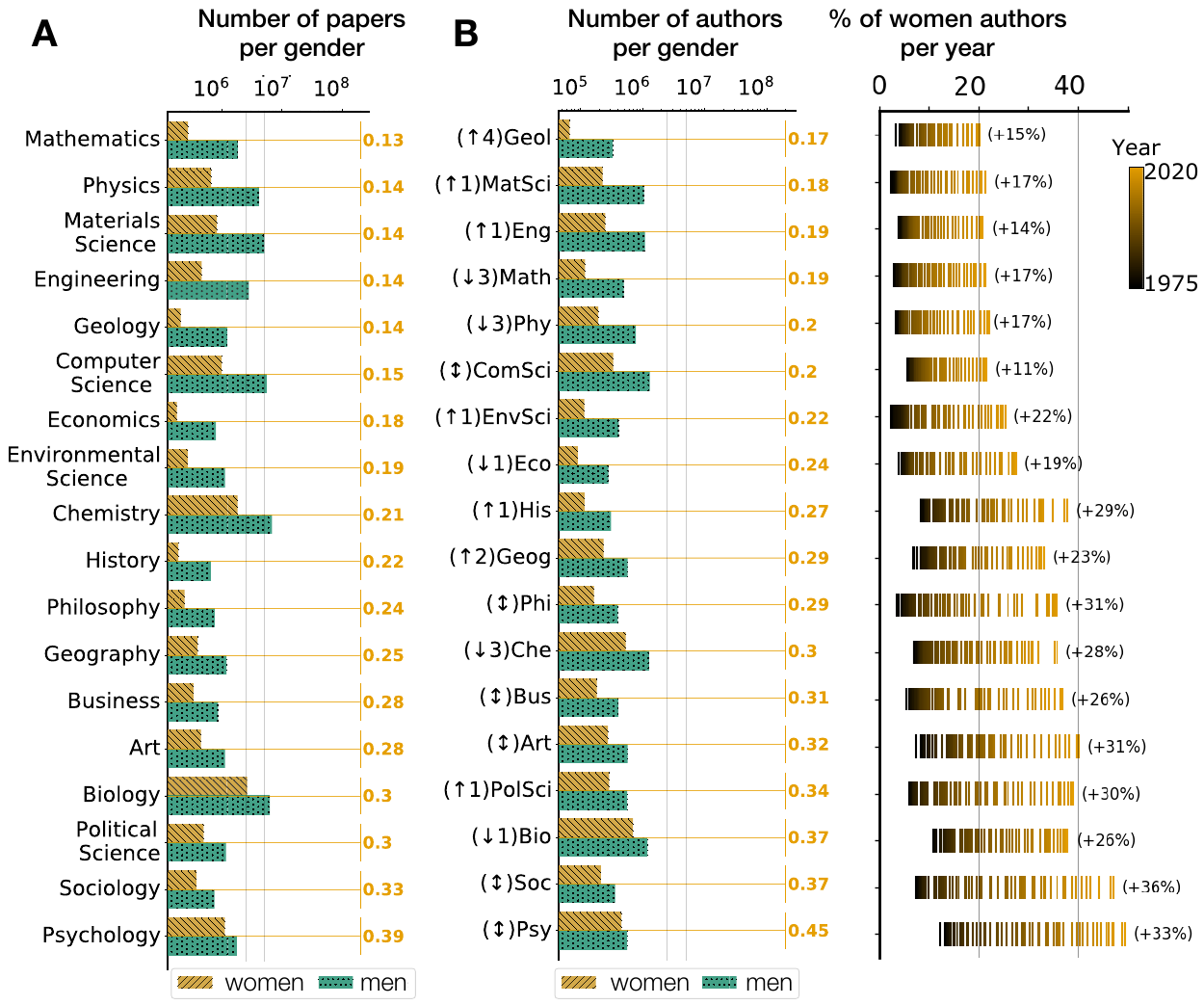}
    \caption{\textbf{Participation of women and men researchers per field.}  Each panel shows the proportion (number) and number (bar chart) of papers in \textbf{A} (authors in \textbf{B}) written by at least one author classified as women or men per field in the Semantic Scholar dataset. All orange (diagonal-lined) bars refer to women, and green (dotted) bars refer to men. In \textbf{A}, the order of the fields goes from top (smallest) to bottom (highest) proportion of papers with at least one author classified as a woman, which is the proportion written to the right of the panel. These proportions are solely for women, and complementary proportions represent papers without women authors. In \textbf{B}, the order corresponds to the lowest to highest proportion of women authors in the field, with arrows comparing both panels: $(\uparrow)$ when the field decreased its position, $(\downarrow)$ increased, and $(\updownarrow)$ maintained the same place, and the number of changed positions when necessary. The right panel in \textbf{B} refers to the growth over time of the percentage of women authors per field, with each line positioned at its respective percentage value and its colour corresponding to the year referred to in the colour bar. The value in parentheses represents the total percentage growth in women from 1975 to 2020. In this panel, we show proportions solely for the number of women, as the proportions of men are complementary. In the \textbf{A} and \textbf{B}-left, the x-axis is on a logarithmic scale to show the smallest values for women in some fields, with reference lines at 2.5 and 5 million to guide the reader. Refer to \figurename~\textbf{S4} for the same bar plots on a linear scale.}
    
    \label{fig:percentage_raw_papers}
\end{figure}

Our dataset shows that women remain a minority in academia across most fields from 1975 to 2020, both in the number of publications and in their share of authorships, with substantial variation across disciplines (\figurename~\ref{fig:percentage_raw_papers}). Specifically, to compare the participation of women across fields, we ordered the fields by the percentage of papers authored by at least one woman, from lowest to highest. This metric includes papers where women appear in any authorship position, as well as papers authored exclusively by women. In our dataset, fields such as Mathematics, Physics, and Materials Science have the lowest proportion of papers written by women (14\%), while Psychology has the highest (39\%) (\figurename~\ref{fig:percentage_raw_papers}\textbf{A}). This ordering grouped fields with similar gender representation; for instance, the pSTEM fields (\eg Mathematics, Physics, Materials Science, Engineering, Geology, and Computer Science) displayed the lowest participation of women. In contrast, the Social Sciences (\eg Political Science, Sociology, and Psychology) had the highest participation. When disaggregating by papers written solely by women or men, or by both genders (\figurename~\textbf{S5}), from History to Psychology, the number of papers written solely by women authors is larger than the number of papers with both genders, but both categories are smaller than the number of papers written solely by men.

When reordering the fields by the percentage of women authors (considering any position in the authorship list), the order of the fields slightly changes(\figurename~\ref{fig:percentage_raw_papers}\textbf{B}-left). For example, Geology shifts from having the fifth-lowest proportion of papers with at least one woman author to having the lowest proportion of women among all authors. This discrepancy shows that women’s representation as authors does not necessarily align with the share of papers authored by women.

The ratio of women in academia has increased over time, possibly indicating the emergence of better opportunities for women to enter the fields~\cite{UnescoWomeninAcademia}. When comparing the percentage of women in each field over time (\figurename~\ref{fig:percentage_raw_papers}\textbf{B}-right), pSTEM fields show the smallest increases (below 20\%), while the Social Sciences exhibit the highest (above 34\%) from 1975 to 2020. Thus, the slower pace of growth in pSTEMs suggests that, despite organisational efforts to increase women's participation in academia~\cite{UnescoWomeninAcademia}, it is consistent with previous documented barriers such as discrimination~\cite{doering2023me}, social norms~\cite{fox2020gender}, and stereotype threats~\cite{van2012slaying,van2022equal,beasley2012they}.

The growth in the percentage of women in each field is concentrated among early-career researchers rather than by retention at senior levels. In our analysis, this growth is largely attributable to researchers who published in a single year, as well as to those in the early stages of their careers (2 to 10 years of publication). In contrast, the percentage of senior women, defined as those with more than 25 years of publications, shows only small increments and remains consistently below 25\%, coinciding with high dropout rates (in \sectionname~\textbf{S3}, we show career stages over time as well as dropout rates). This pattern is consistent with prior work showing that women in Physics~\cite{neuhauser2023improving} and Computer Science~\cite{jadidi2018gender,jaramillo2021reaching} tend to have shorter careers, with lower entry and higher annual dropout rates than men, and suggests that increasing participation has not yet translated into greater retention at senior levels.

\subsection*{Underrepresentation of women in top-ranked positions}
\label{sec:underrepresentation}
 
 We measured three variables, each reflecting different aspects of academic performance and social processes: \textit{i)} productivity, measured by the number of publications, representing the individual research output of a researcher; \textit{ii)} citations, viewed as a collective signal of recognition and appreciation by the scientific community~\cite{Bornmann2014}; and \textit{iii)} coauthorships, measured by the number of unique co-authors, serving as a proxy for social capital~\cite{FronzettiColladon2020,LI2013}, which facilitates success and advancement in academic careers~\cite{Scott2001,Zingg2020}. We then ranked all authors by field according to these three variables and analysed the representation of women and men across these rankings.

\begin{figure}[!ht]
    \centering
\includegraphics[width=0.9\textwidth]{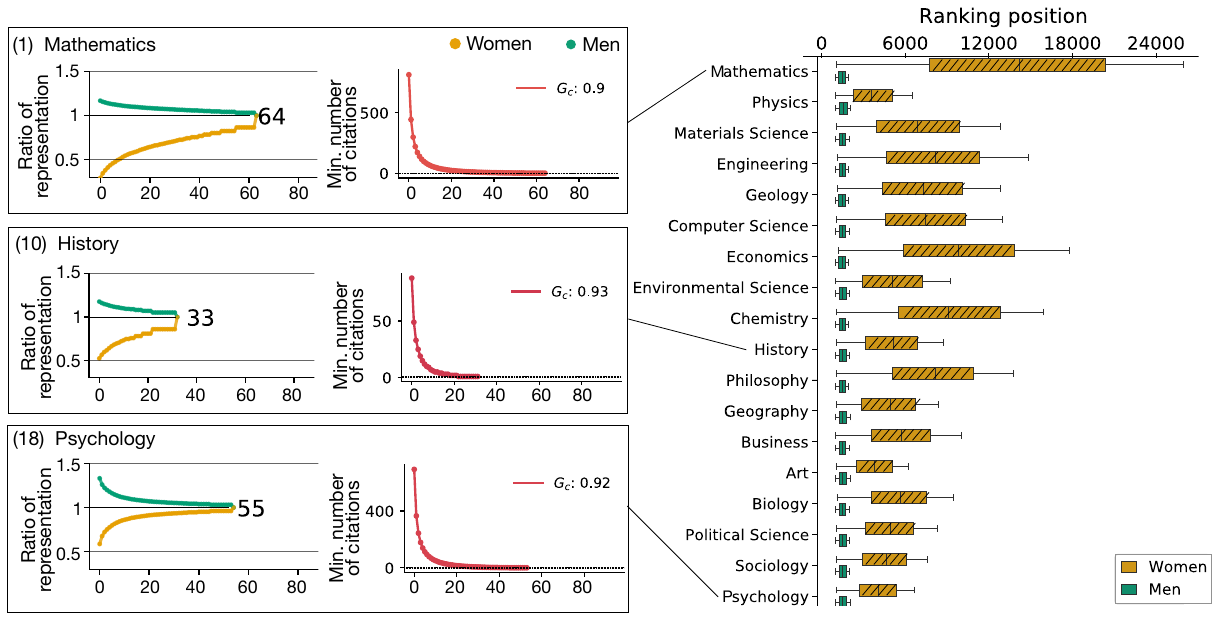}
    \caption{\textbf{Representation of women and men in top-ranked positions.} \textbf{Left panels}: in the left subpanels, the $y$-axis refers to the representation of each gender in the $k\%$ top-ranked position, orange for women and green for men. $y\approx1$ is a fair representation, $y>1$ is overrepresentation (in most cases for men), and $y<1$ is under-representation (in most cases for women) -- lines at 1.5, 1, and 0.5 are to guide the reader. The representation is the division between the proportion of one gender in the top-ranked $k\%$ of citations and their total participation in the field (orange for women and green for men) as shown in \equationautorefname~\ref{eq:represenation}.  The line in 1 finishes at the percentile where women and men reach ``fair'' representation, resulting in the same proportion in the top-ranked position and participation in the field: 64 in Mathematics, 33 in History, and 55 in Psychology. The right subpanels correspond to the minimum number of citations necessary to be in the top-ranked $k\%$, and $G_{metric}$ represents the Gini coefficient for the distribution of citations. Both subpanels show results for fields with the lowest (1), medium (10), and highest (18) participation of women. \textbf{Right} panel corresponds to the position of the top $1{,}000$ most cited women and men per field. We ranked all researchers by citations, and the box plots show the variance in their positions in the ranking (for rankings of the other two metrics and for normalised versions of this ranking to the top 1\% of authors, see \sectionname~\textbf{S4}).
    }
    \label{fig:Representation_top_ranked}
\end{figure}

 The rankings reveal a significant imbalance between men and women authors across all fields. To illustrate this disparity, we measured how changing the percentile of researchers considered in the top-ranked positions relates to the representation of women and men in these positions, further details in Methods and in \equationautorefname~\ref{eq:represenation}. As shown in \figurename~\ref{fig:Representation_top_ranked} (left subpanels), when accounting for accumulated citations up to 2020, women were highly underrepresented within the highest-ranking positions (\eg the top 5th percentile with values below 0.5). We found that fields with higher women's participation reach fair representation ($\approx 1$) in a smoother way (\eg Psychology gradually reaching fair representation at the 55th percentile). In contrast, fields with low participation of women reach fair representation more abruptly (\eg for Mathematics at the 64th percentile, the entire population is reached while in the 63rd percentile women are still highly underrepresented). Our complete analyses for all fields using the three metrics are provided in \figurename~\textbf{S17}. 

While all fields are highly unequal in terms of citations (\figurename~\ref{fig:Representation_top_ranked}-right subpanels), measured by the Gini coefficient ($G_c$ in the figure), the underrepresentation of women in top-ranked positions is more pronounced in fields with a more unequal distribution of productivity (pSTEMs shown in \figurename~\textbf{S18}-left column). In the most unequal fields, researchers require greater productivity and more citations to reach top-ranked positions, and fair gender representation is achieved only at higher percentile ranks. For instance, to be among the top $1\%$ of most-cited researchers in Mathematics, a researcher needs around $1{,}000$ citations and at least $30$ papers. Within this top $1\%$, women account for approximately $5.6\%$, less than half of their overall participation in the field, where $16\%$ of the researchers are women. Fair representation of women in Mathematics (\ie a ratio of $1$) is only achieved at the 64th percentile of the ranking, at which point the full $16\%$ of women in the field are included. In contrast, researchers in History need around $100$ citations to be in the top $1\%$, with fair representation of women occurring earlier, at the 33rd percentile (\figurename~\textbf{S18}).

Across all ranking positions and the three metrics, citations are the most unequally distributed metric, consistent with a rich-get-richer multiplicative process (\sectionname~\textbf{S4}~middle column). Regarding productivity, this metric is the least unequal but most closely related to the order of women's participation across fields (Higher $G_c$ for fields with fewer women). In addition, productivity is also the fastest metric to reach fair representation of women in top-ranked positions across all fields, and this could be related to its lower inequality, as most researchers have at least one paper (\figurename~\textbf{S17}~and~\textbf{S18}~left panel). At last, the number of coauthorships has medium values of inequality, although in Sociology and Philosophy, women have slightly more representation than men after the 5th percentile (\figurename~\textbf{S17}~and~\textbf{S18}~right panel)

 To complement this ranking analysis, we focus on the position and fairness of a fixed set of top-ranked researchers. In \figurename~\ref{fig:Representation_top_ranked}-right, we present the ranking positions of the top $1{,}000$ most cited women and the $1{,}000$ most cited men in each field. A significant difference is observed between the $1{,}000$ top-ranked men and women, with the largest gap found in Mathematics, measured by the exposure ratio~\cite{Singh_2018}, where top-ranked women rank 24 times farther down than top-ranked men. Similar trends are observed in terms of productivity and coauthorships: in pSTEM fields, women occupy positions much farther down the rankings (12 times lower than men on average), while in the Social Sciences, women's positions, though still lower (6 times farther than men on average), are closer to parity compared to pSTEM fields (\figurename~\textbf{S12} and \figurename~\textbf{S13}). We repeated this analysis for the 1\% top-ranked women and men per field in \sectionname~\textbf{S4} by normalising the ranking position based on the field size. We found that the distance between the rankings decreases linearly as the group sizes of women and men within each field become more balanced.
 
\subsection*{Rigidity in rankings over time}
\label{sec:dynamic_rankings_over_time}

As women's participation in academia increases, we examine the rate at which women are entering top-ranked positions in citation counts. An open ranking system would be composed of rankings that are more flexible and movable (\ie new people are frequently introduced to the rankings), whereas a rigid ranking system suggests minimal introduction. In the academic context, previous studies have shown that university rankings tend to exhibit a rigid structure, with limited turnover in the top positions~\cite{Iniguez2022} and the presence of super-stable individuals in top-ranked positions within citation networks~\cite{Ghoshal2011-ux}.

\begin{figure}[!th]
    \centering
\includegraphics[width=\textwidth]{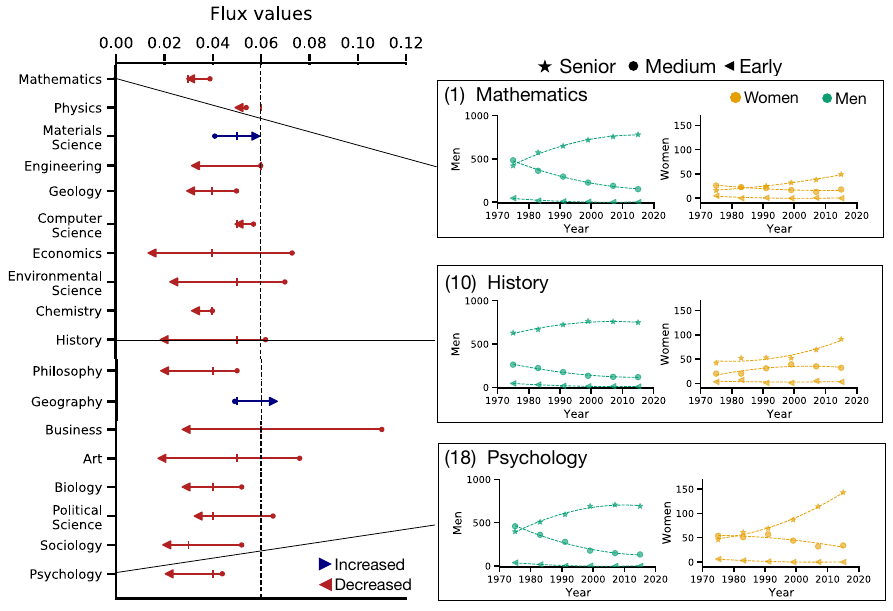}
    \caption{\textbf{Change of the rankings over time.} On the left side of the plots, we display the flux values of the top $1{,}000$ most cited researchers in each field across years. Each line starts at the flux value in 1976 represented with a circle (first year, $\mdsmblkcircle$). The final flux value in 2020 is represented with a triangle to the left (decreased flux, $\smallblacktriangleleft$) and the right (increased flux, $\smallblacktriangleright$), and the horizontal line is the average value of the 45 years (average, \textbf{\textbar}). Fields with decreasing flux values are shown in red, while those with increasing flux values are in dark blue. As in previous plots, fields are ordered by the participation of women from lowest (top) to highest (bottom), and a vertical dashed line at 0.06 is a baseline for the reader. On the right side, detailed visualisations showing the number of researchers, categorised by gender, are presented for different career stages per year for the fields with the lowest participation of women (Mathematics), medium participation (History), and the highest participation (Psychology). These curves are derived from polynomial regressions, with further details provided in \sectionname~\textbf{S5}.}
    \label{fig:Temporal_rankings}
\end{figure}

We measured ranking rigidity using the flux value, defined as the percentage of researchers who enter or exit the ranking in a given year $(y)$ relative to the previous year $(y-1)$, with values ranging from 0 (most rigid) to 1 (least rigid)~\cite{Iniguez2022}. We then analyse the flux among the top $1{,}000$ researchers in each field, ranked by citation counts, to examine how rigidity evolved over time (\figurename~\ref{fig:Temporal_rankings}). Our findings reveal that most fields have relatively low flux values (below 0.12), with decreasing flux values over time observed in most fields except for Materials Science and Geography (\figurename~\ref{fig:Temporal_rankings}-left, and for all years and fields in \sectionname~\textbf{S5}). 
Specifically, Business, Art, Economics, Environmental Science, History, and Political Science initially showed lower ranking rigidity but saw an increase in rigidity from 1976 to 2020. In contrast, pSTEM fields showed no substantial changes in ranking rigidity over time. 

We perform a robustness test computing the flux values for different rankings of fixed size: 100, 500, $2{,}000$, and $5{,}000$~(\figurename~\textbf{S19}). The results remain similar as the ranking becomes more rigid over time, with values lower than 0.12. A few fields, such as Physics, show small increases in flux value when considering the top 100 ranking. Still, for the $5{,}000$ top-ranked researchers, all flux values become more rigid. In addition, considering that each field has different sizes and that the fields' growth could bias the ranking flux, we also computed dynamic flux values considering the top 1\%, 5\%, and 10\% of researchers per field (\figurename~\textbf{S20}). In this case, we retrieved a vector of length $n$ containing the top-$k\%$ ranked researchers for year $y$ and compared its flux value with that of the top-$n$ researchers in year $y-1$. 

To understand the gender differences in the flux values, we analyse the representation of women in the top $1$st percentile for citations over time in \sectionname~\textbf{S5.2} (a temporal version of the analyses conducted in the previous section). In all fields, the representation of women was generally $0.5$ or lower in 1975. Although it slowly increased to values above $0.5$ in most fields, some fields witnessed a decrease or stagnation. Specifically, the representation of women remained below or equal to $0.5$ in Mathematics, Geology, Engineering, Computer Science, Economics, Chemistry, History, Philosophy, Biology, and Business, which are fields that also showed no change or declining trends in the proportion of new women entering top-ranked positions (\figurename~\textbf{S22}). 

Finally, we analyse the proportions of women and men across career stages among the $1{,}000$ most-cited researchers over time (\figurename~\ref{fig:Temporal_rankings}, right subpanels). In all fields, the participation of senior men shows a logarithmic increase, while the participation of men at the medium career stage declines. In contrast, there is an exponential increase in the participation of senior women, while their values remain five or more times smaller than those for men, even in the Social Sciences, where women's overall participation is higher. As expected, early-career researchers have almost no participation in the $1{,}000$ most cited positions across all fields (\figurename~\textbf{S21}).

In \sectionname~\textbf{S5.4}, we replicate our ranking analyses with thresholds above 0.8 (0.9 and 0.95) for the gender inference; the threshold of 0.8 places women in lower rankings when separating them by gender in the strict sample of 1,000 researchers per gender, but shows no difference when considering the field size with the top 1\%. Due to the consistency when considering the sample size, and that this smaller threshold of 0.8 permits including more women in the top positions overall, especially senior women researchers, we keep the main manuscript results with the 0.8 threshold.

In summary, when we fix the number of top-ranked researchers, we observe low flux values and a slow increase in the proportion of senior women entering the rankings. When considering the top $1\%$ of researchers in each field, annual growth allowed some newcomers to enter the rankings. However, whilst the proportion of senior women in the rankings increased, it remained lower than their overall representation in the fields. Consequently, the most notable changes in the rankings occurred not at the very top positions but at intermediate ranks, situated between the top $1{,}000$ and the top $1\%$ of researchers (\eg the top $20{,}000$ positions for Physics and the top $10{,}000$ positions for Sociology). We note that the flux measure is computed over all researchers in a ranking and is not gender-specific: it documents decreasing turnover, and the gendered composition of these top-ranked positions, but our analysis does not test whether rigidity contributes to the gender differences we observe.

\subsection*{Gender gaps in academic recognition}
\label{sec:regressions}

The previous analyses show that women remain underrepresented in top-ranked positions and that citation rankings tend to be stable and rigid over time. However, these aggregate patterns alone do not show whether citation disparities mainly reflect differences in participation, prestige, and career composition, or if they remain among researchers with similar publication profiles. Direct comparisons are difficult, as fields differ in their gender composition, publication practices, authorship conventions, and career length. In addition, due to accumulated advantage, men have historically been more numerous and more visible in academia~\cite{lariviere2013bibliometrics}.

To assess whether citation disparities persist among researchers with comparable publication profiles, we focus on highly productive researchers within each field and career stage and construct matched samples. These profiles combine career age and productivity, disaggregated by authorship position and venue citation quartiles (Q1 to Q4, from highest- to lowest-ranked publication venues), which served as a proxy for venue prestige. Because no single comparison can remove all sources of heterogeneity, we employ three complementary matching strategies: \textit{i)} \matchingdistribution\ to preserve the observed gender composition of highly productive authors within each field and career stage, providing a baseline view of citation gaps among researchers who already occupied these positions. \textit{ii)} \matchingmen\ to anchor the comparison on highly productive men and identify the closest available women counterparts based on observed publication profiles. \textit{iii)} \matchingwomen\ to reverse the comparison by anchoring on highly productive women and identifying the closest available male counterparts. (see Methods and \sectionname~\textbf{S6} for details). 

We consider these strategies as complementary descriptive contrasts each anchored on different reference groups, rather than interchangeable estimates of a single underlying gender effect. All three are conditional on entry into the highly productive group. Because reaching the top $k\%$ of productivity may itself be shaped by gender-related career conditions, including the shorter careers and higher dropout rates documented above, the women in these samples are a selected group. The estimates should therefore be read as conditional associations among highly productive researchers rather than as the overall gender gap in citations within a field. Within this selected group of researchers, comparing the three strategies allow us to examine their citation disparities. If comparable publication profiles translate into comparable citation outcomes for women and men, estimates from these strategies should converge. When they do not, differences across estimates indicate how gender-associated differences in expected citations change with the reference group, while differences in match quality reveal whether comparable researchers of the other gender are available in the same region of the observed publication-profile space.

We then examine citation outcomes among the top 5\% and 10\% most productive researchers between 1990 and 2020, when women's participation is sufficiently large for significant comparisons. For each matching strategy, we estimate Negative Binomial fixed-effects models separately by field and in a pooled specification including all fields at once, suited to skewed and overdispersed citation counts. The models adjust for gender, career stage group (early, medium or senior), cohort (decade of first publication in the field), number of coauthorships, and  productivity disaggregated into four venue prestige categories and three authorship positions, productivity--coauthorship interactions, year fixed effects, and, in the pooled model, field fixed effects. Full model details and diagnostics are provided in Methods and \sectionname~\textbf{S6}.

\begin{figure}[!h]
    \centering
    \includegraphics[width=\linewidth]{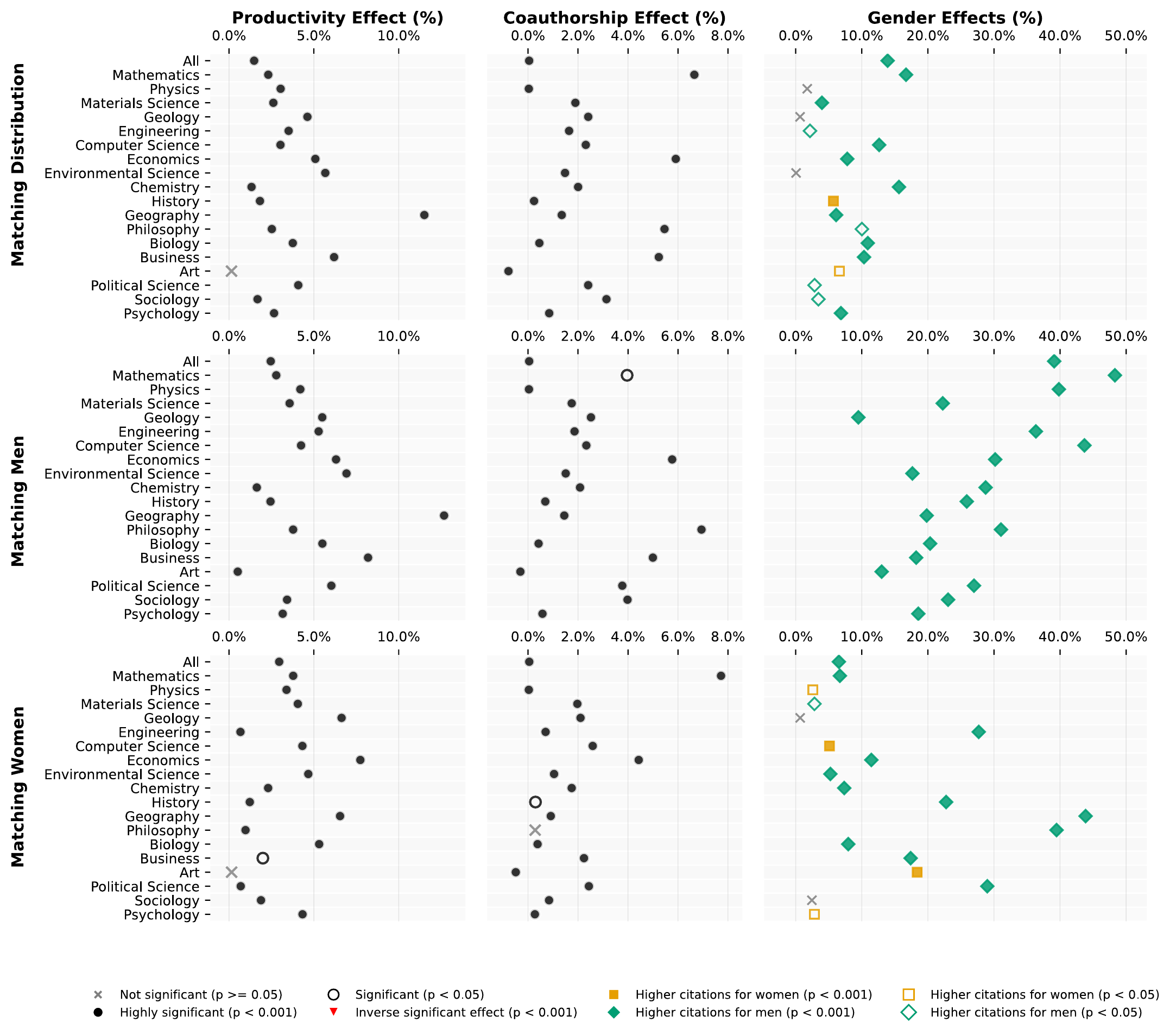}
    \caption{\textbf{Significant associations of productivity as first authors in Q1 venues, coauthorships, and gender with citations across fields for similar top 5\% authors.} Each row represents a different matching strategy: top row for~\matchingdistribution; middle row for~\matchingmen; and bottom row for~\matchingwomen. 
    The first and second columns show the percentage change in expected citations per one additional paper in \textit{Productivity} as first author in a Q1 venue and per one additional \textit{Coauthorship}, respectively.
    The third column shows the absolute magnitude of the estimated percentage difference in expected citations for women relative to men, with orange $\smblksquare$ indicating the \% of \textit{more} expected citations for women and green {\fontsize{6.5}{8}$\blacklozenge$} indicating the \% of \textit{less} expected citations for women compared with men.
    Statistical significance is shown through marker fill: filled markers indicate $p<0.001$, open markers indicate $0.001 \leq p<0.05$, and grey crosses indicate non-significant estimates.
    For example, in the first row of \matchingdistribution\ for `All' (pooled model across fields), one additional first-author Q1 paper is associated with nearly 1.5\% more expected citations, one additional \textit{Coauthorship} with 0.04\% more expected citations, and women have 13.9\% less expected citations than men (filled green diamond); all three estimates are significant at $p<0.001$.
   }
    \label{fig:gender_disparities_regressions_5pct}
\end{figure}

Results are summarised in \figurename~\ref{fig:gender_disparities_regressions_5pct} for the top $5\%$ most productive researchers. In the first column, for productivity, the figure reports the transformed model coefficient for first-author publications in Q1 venues, expressed as a percentage change in expected citations. We reported this category because Q1 publication venues account for the largest share of publications among top-ranked researchers, and first-author Q1 publications are the most frequent in 9 fields (\figurename~\textbf{S32}). The second column reports the corresponding percentage change associated with coauthorships, and the third column reports the estimated citation gap between women and men. Corresponding summary results for the top $10\%$ are shown in \figurename~\textbf{S40}, with complete regression tables for the pooled models and each field-specific model reported in Tables~\textbf{S22}-~\textbf{S59}.

Across comparison strategies, gender-associated citation gaps are substantial and favour men in most fields. In the pooled estimates, women receive 13.9\% fewer expected citations than men under \matchingdistribution, 39.1\% fewer under \matchingmen, and 6.5\% fewer under \matchingwomen. On the log-model scale, these gaps are comparable to the expected citation gains associated with approximately 10.2, 20.4, and 2.3 additional first-author Q1 publications, respectively.

The differences in magnitude of the gender-associated citation gaps across strategies are consistent with how closely researchers could be matched. We quantify profile similarity with the Mahalanobis distance. Distances are larger in \matchingmen\ than in \matchingwomen: women paired to highly productive men are, on average, farther from their men counterparts (\figurename~\textbf{S34}) than men paired to highly productive women are from their women counterparts (\figurename~\textbf{S35}), full analyses in \sectionname~\textbf{S6.2}. Post-matching balance diagnostics show the same pattern: \matchingwomen\ achieves close balance across the 13 matching variables, whereas \matchingmen\ retains larger residual differences (\tablename~\textbf{S7}). Therefore, \matchingwomen\ produces closer publication-profile matches, whereas \matchingmen\ captures a men-anchored upper-tail contrast in which the nearest women counterparts are, on average, more distant. In contrast, \matchingdistribution\ preserves the observed separation between highly productive women and men in conventional top-percentage rankings. This asymmetry reflects limited common support in the upper tail: for some highly productive men, no woman with a closely comparable publication profile exists in the same field and career stage, so the nearest available counterparts are, on average, younger and have fewer Q1 publications. We therefore do not interpret the \matchingmen\ estimate as a fully like-with-like contrast between comparable researchers. Rather, it quantifies the regression-adjusted citation difference between the men who define the upper tail and the closest women available to them, which is itself an informative feature of these fields rather than a limitation of the method to be corrected.

Looking at each comparison strategy in more detail, the \matchingdistribution\ models (top row of \figurename~\ref{fig:gender_disparities_regressions_5pct}) provide the most moderate differences: field-level citation gaps have the smallest average magnitude and are more homogeneous across fields, although their direction and statistical support remain field-specific. Women received significantly less expected citations in 13 of 18 fields ($p<0.05$; 9 at $p<0.001$). The largest differences appeared in Mathematics ($16.7\%$), Chemistry ($15.6\%$), and Computer Science ($12.6\%$). The pattern reversed in History and Art, where women received $5.7\%$ and $6.6\%$ more expected citations, respectively. Differences were not statistically significant in Environmental Science, Geology, and Physics. Taken alone, this baseline could suggest a more moderate pattern of inequality. However, like the two matched comparisons that follow, it reflects the observed highly productive population, where women who reach the top may already differ from the broader population of women researchers.

The \matchingmen\ models (middle row of \figurename~\ref{fig:gender_disparities_regressions_5pct}) anchored the comparison on men in the top $5\%$ of productivity. In this comparison, estimated gaps favoured men in every field and exceeded the corresponding distribution-preserving estimates, ranging from $9.5\%$ in Geology to $48.3\%$ in Mathematics. The pattern broadly aligned with women's participation across fields: the largest differences appeared in lower-participation pSTEM fields (except Materials Science and Geology), while the smallest gaps appeared in fields with higher women's participation, including the Social Sciences and Art.
To check that this result was not driven by the most distant matches, we repeat the \matchingmen\ analysis while progressively excluding matched pairs with the largest Mahalanobis distances, up to the most distant 20\%. The pooled gap decreased from 39.1\% to 34.9\%, but remained in the same direction and statistically significant. The mean Mahalanobis distance fell from 3.03 to 1.93, indicating that the retained pairs were closer overall in their publication profiles, while mean $|\mathrm{SMD}|$ (which captures differences between the matched groups on the individual matching variables) increased from 0.304 to 0.333. Thus, although trimming improves the overall proximity of the matched pairs, a sizeable citation gap and substantial differences in observed publication profiles remain, indicating that the gap extends beyond the most distant matches but remains associated with the lower comparability between highly productive men and their closest women counterparts (Section~\textbf{S6.8}; Table~\textbf{S60}).

The \matchingwomen\ models (bottom row of \figurename~\ref{fig:gender_disparities_regressions_5pct}) provided a complementary comparison, anchoring the analysis on women in the top $5\%$ of productivity and identifying their closest men counterparts.
Overall, citation gaps narrowed in some fields but increased in others: women received significantly fewer expected citations in 12 fields ($p<0.05$; 11 at $p<0.001$), and in 9 of these fields the gap was smaller than in \matchingmen. Differences were not statistically significant in Geology and Sociology. Gaps favoured women in Art ($18.4\%$, $p<0.001$), Computer Science ($5.1\%$, $p<0.001$), Psychology ($2.8\%$, $p<0.05$), and Physics ($2.6\%$, $p<0.05$). At the same time, Geography ($43.9\%$, $p<0.001$), Philosophy ($39.5\%$, $p<0.001$), and Political Science ($29.0\%$, $p<0.001$) remained fields where gaps favoured men and were even larger than the corresponding estimates from \matchingmen.

This field dependence is consistent with cross-field differences in the alignment between productivity and citation rankings (\figurename~\textbf{S43}). Where this alignment is weaker, similar publication profiles may translate less directly into citation recognition. In addition, our results remain consistent when we remove self-citations from the dependent variable and re-run all models using only citations made by other researchers (see \sectionname~\textbf{S6.6} for further details). Therefore, self-citations do not qualitatively change the estimated gender effects among top matched researchers. The estimates are also robust to a stricter gender-inference threshold: re-estimating all three matching strategies on authors whose gender was inferred with at least 0.9 probability leaves the direction and significance of the gender-associated gaps unchanged, with \matchingdistribution\ reversing in History and Geology (\sectionname~\textbf{S6.7}).

Together, our results show persistent asymmetries in academic recognition. In the distribution-preserving comparison, citation gaps were smaller and more field-specific. When the comparison was anchored on highly productive men, estimated citation gaps favoured men across all fields, and the nearest women counterparts were, on average, more distant. When the comparison was anchored on highly productive women, gaps narrowed in some cases and sometimes disappeared or reversed, but this did not imply parity at the upper tail, because the matched men were not necessarily the most productive men in their field. Even in comparisons that yielded the closest observed matches, similar publication profiles did not always translate into similar citation recognition, revealing a consistent asymmetry in which men generally received more citations than women among highly productive researchers.

\section*{Discussion}
\label{discussion}

The increase in women's participation in academia over the past decades is likely influenced by educational policies promoting gender equality from national and international organisations such as the United Nations~\cite{Huck2023}. This rise in participation has positively influenced the recognition of underrepresented communities of scientists~\cite{Huang2020}. However, various studies continue to show the significant underrepresentation of women in top-ranked positions, especially in pSTEM fields~\cite{jaramillo2021reaching,lariviere2013bibliometrics,macedo2023academic} and in decision-making and leadership roles~\cite{Yousaf2017}. Women's limited presence in top-ranked positions may be linked to historical and cumulative disadvantages, as well as the lack of social support for pursuing research careers~\cite{van2022equal}. Nevertheless, studies exploring the impact of increased women's participation and representation in top-ranked positions remain limited, particularly in terms of the rigidity of rankings, productivity, coauthorship networks, and temporal dynamics. In this study, we studied this gap by estimating gender differences in \textit{i)} the representation of top-ranked positions, \textit{ii)} the evolution and rigidity of academic rankings over time, and \textit{iii)} the effect of productivity, coauthorships, and gender in gaining citations.

This study examines gender differences in coauthorship networks and ranking patterns from a cleaned dataset of more than 60 million papers published between 1975 and 2020, retrieved from the Semantic Scholar Open Research Corpus dataset~\cite{Lo2020}. First, we infer the gender of the authors in each field based on their first and last names (\sectionname~\textbf{S1}), and we organise the fields by women's participation, ranging from the lowest (Mathematics) to the highest (Psychology). We then compute gender representation, measured as the ratio of the proportion of women in the top-$k\%$ ranking over the proportion of women in the entire field, for different $k$ percentiles, across three metrics: productivity (number of papers), citations, and coauthors degree (number of coauthors). Next, we analyse women's representation relative to their participation in each field and the temporal evolution of these rankings. Finally, within each field and career stage, we compare highly productive researchers with similar publication profiles to assess their similarity in citation gains.

Our results indicate that, by group size, women are a minority in all fields. When ordering the fields from lowest to highest participation, they align by disciplinary categories, such as pSTEM fields, Life Sciences, and Social Sciences. When analysing representation in top-ranked positions, we found that women tend to be underrepresented in all fields for rankings up to the 10th percentile, with representation increasing from 0.5 (half of their overall participation in the field) to a more equitable representation as we move to lower-ranked positions. These increments are slower in fields with the lowest women's participation (pSTEM fields), and they only reach fair representation when considering the entire field (with similar proportions in top-ranked positions as in the overall population). In comparison, the increments are faster and smoother in fields with the highest women's participation (Social Sciences). These contrasting patterns between pSTEM and Social Sciences highlight that women’s advancement is not solely determined by their overall participation in a field. This finding is interesting because even fields with higher women's participation exhibit low representation in top-ranked positions~\cite{van2022equal}. Yet, it aligns with previous evidence showing that women remain underrepresented in positions of power and prestige, shaped by academic inequality regimes, entrenched labour and power divisions, gender-based discrimination, shorter career spans, the `leaky pipeline', and persistent social constraints that limit women's career advancement~\cite{acker2006inequality,fox2020gender,doering2023me,wang2021science}.

In addition to the descriptive analyses, we estimated Negative Binomial fixed-effects models across three complementary matched samples of researchers to assess the associations of productivity, coauthorship, and gender with citation gains, while controlling for venue prestige, authorship position, cohort, career age and year. Our results show that gender is significantly associated with citation gains and that gaps persist among highly productive researchers with similar publication profiles, although their magnitude varies across fields and comparison strategies. In the distribution-preserving comparison, gaps tend to be smaller and more field-specific. When the comparison is anchored on highly productive men, they have farther distances to the comparable women and the citation gaps favour men in all fields, with the largest gaps in disciplines where women’s participation is lower (pSTEM). When the comparison is anchored on highly productive women, gaps attenuate in some fields and in a few disappear or reverse (\eg Art and Computer Science), while other fields show high gaps favouring men (ordered from largest to smallest: Geography, Philosophy, Political Science, Engineering, and History). Taken together, these comparisons show that similar publication profiles do not always translate into similar gained citations, self-citations do not play a major role and even when correcting for several variables, the upper tail of citation distributions is still highly composed of men.

At the time of publication, our work is based on one of the largest datasets of researchers with inferred gender information, including 4,553,912 women and 11,956,727 men across various disciplines, and it also accounts for authorship position and a proxy for publication venue prestige. Nevertheless, our work has methodological constraints, including the inability to detect non-binary or transgender individuals, potential biases related to country or language, variations in naming conventions and typos, and the limitations of assigning gender based solely on names~\cite{Gonzalez-Salmon2024-hz}. Researchers whose gender could not be inferred with at least a probability of 0.8 as either women or men from their first or complete names (if discrepancies of gender assignation in different countries exist) were excluded from the analyses (see \sectionname~\textbf{S1}, \sectionname~\textbf{S5.4} and \sectionname~\textbf{S6.7} for robustness checks at stricter thresholds). This exclusion is not random across fields: name-based inference resolves fewer names in fields with lower women's participation (\figurename~\textbf{S2}), so coverage is lowest precisely where the gaps we estimate are largest. Stricter probability thresholds do not address this, since the issue is differential coverage of names by linguistic and geographic origin rather than classification confidence alone. Our estimates should therefore be understood as describing the subpopulation of researchers whose names are resolvable by these tools.
A further constraint concerns the scope of the matched comparisons: our regression estimates are conditional on researchers having already reached the top 5\% or 10\% of productivity within their field and career stage. They therefore characterise citation differences within this selected group and do not speak to gender disparities in reaching it, which our descriptive analyses suggest are themselves substantial.

Addressing these limitations and incorporating more inclusive gender detection methods remain important directions for future work. Beyond gender inference, our large-scale analysis is also constrained by the coverage and metadata available in Semantic Scholar. In particular, Semantic Scholar has limited coverage of books and monographs, which may be especially relevant for fields in which these formats are central research outputs, and our data do not include institutional affiliation or geographic information. Additionally, our large-scale analysis has excluded other factors, such as gender differences in self-promotion on social media platforms~\cite{peng2023gender,spagert2025doing}, gendered citation practices~\cite{fox2020gender}, and inequalities in international coauthorships~\cite{momeni2022many}. A systematic analysis of these factors using mixed methods, including experts from each of the 18 studied fields, remains an open area for future work.

Our results have implications for scholarly research engines, such as Google Scholar, which rank researchers and papers based on the number of citations they receive. We document that citation rankings became more rigid over time, but our analysis does not identify the mechanisms behind this trend. Prior work suggests candidate explanations: cumulative visibility bias in search and recommendation systems, described as a `rich get richer' effect~\cite{farber2023biases}, which has been shown to affect the visibility of new researchers and minorities~\cite{Espín-Noboa2022}; and multiplicative processes over skewed distributions, which become more rigid over time~\cite{Ghoshal2011-ux} and yield low mobility at the highest ranks~\cite{Sun2023}. Whether increasing rigidity contributes to the underrepresentation of women we document, or simply coincides with it, remains an open question. Answering it would require designs that model entry into and exit from top-ranked positions by gender, together with barriers such as affiliation improvement~\cite{macedo2023academic} and their possible downstream effects on career advancement and workforce participation~\cite{Gorman2005-ua}.

These findings offer valuable insights for developing evidence-based policies that enhance women’s participation and career advancement across academic fields. While citation and publication counts provide a useful metric of impact, relying solely on them limits our understanding of other forms of representation in academia. Our work reveals structural disparities rather than direct causal gender discrimination; however, based on prior literature, these disparities arise from the complex interplay of hidden variables that affect women’s participation in science~\cite{Cocks1988,Rosenfeld1982,acker2006inequality,Lavere020380,doering2023me,fox2020gender,van2012slaying}. Enhancing the participation of minorities in academia requires alignment with feminist efforts to transform the organisational practices of science~\cite{schiebinger2000has} by:
\textit{i)} expanding training and opportunities for women;
\textit{ii)} strengthening the role of diversity and equity offices in improving childcare and workplace safety; and
\textit{iii)} encouraging men researchers to dismantle cycles of discrimination and share caregiving responsibilities~\cite{llorens2021gender}; and \textit{iv)} understanding that citation-based ranking systems affect online visibility and should fairly represent women and marginalised communities~\cite{Vásárhelyi2021,vásárhelyi2023benefits, peng2023gender,spagert2025doing}. 

While our work focuses on the gender dimension of citation rankings inequalities, one of our main limitations is not accounting for other socio-demographic dimensions that could penalise other historically underrepresented communities in science, such as those from middle and low-income countries~\cite{Gomez2022,Park2023,lerman2022gendered}, ethnic minority groups~\cite{patton2004reflections,Liu2010RACE}, and non-Western epistemologies~\cite{Morgan2018}. Future research should adopt an intersectional perspective by examining not only gender but also other identities, such geographical location, and their interactions with academic productivity and recognition, in order to deepen our understanding of a multifaceted society~\cite{Fox_Tree2022-jn}.

\newpage
\section*{Methods}
\label{sec:Methods}

\subsection*{Data}
We used the dataset from the Semantic Scholar Open Research Corpus~\cite{Lo2020} because its broad disciplinary and temporal coverage enabled systematic comparisons across fields and time. It comprises 19 distinct fields previously defined by the Microsoft Academic Graph project and subsequently used by Semantic Scholar. Whilst the full corpus contains more than 200 million papers, our filtered dataset comprised $119{,}206{,}753$ papers spanning 18 fields (excluding medicine) from 1975 to 2020 (45 years). After inferring authors' genders, the final dataset contained $67{,}696{,}061$ papers. To infer gender, we used Genderize~\cite{genderize}, which assigns gender based on the author's first name. For names used by both genders across different countries (based on the `Gender World Bank' database~\cite{worldbankgenderdictonary}), we searched for the first and last name in Namsor~\cite{Namsor}. For both Genderize and the Namsor full-name check, we retained only assignments to women or men with a probability of at least 0.8 (for both libraries) following previous literature~\cite{Karimi2016}; names below this threshold or otherwise unresolved were excluded from the analyses. Further details on data cleaning, gender inference, threshold checks at 0.9 and 0.95, and dataset description are provided in \sectionname~\textbf{S1}. Our results could also be replicated using more specific topics within fields (e.g. Artificial Intelligence, Network Science, Computational Social Science) and other datasets, such as SciSciNet~\cite{lin2023sciscinet}, OpenAlex~\cite{priem2022openalex}, and ACM Digital Library~\cite{Divakarmurthy2012}, which vary in coverage and quality~\cite{wilinski2026science,gusenbauer2024beyond,Lo2020,haupka2026analysis}.

\subsection*{Top-ranking representation}
To analyse the representation of each gender in the top-ranked positions, we defined a $k$ percentile threshold, where the minimum value of the metric required to be considered in the top $k\%$ was determined from the data. For example, for the top $1\%$ of most-cited researchers, the minimum value required to be in the ranking was $1{,}000$ citations; we then included all researchers in that field with at least $1{,}000$ citations. We computed $\text{P-top}^k_g$ as the proportion of researchers of gender $g$ among the top $k\%$-ranked positions, and $\text{P}_g$ as the proportion of gender $g$ in the entire field. The representation displayed in \figurename~\ref{fig:Representation_top_ranked} and \figurename~\textbf{S17} was then computed as in \equationautorefname~\ref{eq:represenation}:

\begin{equation}
\text{Representation}_g^k=\frac{\text{P-top}^k_g}{\text{P}_g}
\label{eq:represenation}
\end{equation}    

\subsection*{Flux in fixed-size rankings}
To compute the flux value of a ranking, we used the method proposed by I{\~{n}}iguez \etal~\cite{Iniguez2022}, which compares two rankings of the same size and measures their differences from one year to the previous year. We measured the flux by counting the number of researchers who entered or exited the ranking from one year to the next, divided by the fixed size of the ranking. This approach considers only whether a researcher is included in the ranking or not, rather than how many positions they moved up or down. To demonstrate the robustness of our results, we examined flux values for fixed ranking sizes of 100, 500, $1{,}000$, $2{,}000$, and $5{,}000$ across all years. Additional results are provided in \sectionname~\textbf{S5.2}.

\subsection*{Flux in dynamic-size rankings}
To account for the size difference between fields, we also analyse the flux in the rankings for different percentiles of the top-ranked positions. To account for the fact that the percentiles in different years can alter the number of individuals, we computed a dynamic version to calculate the flux values. We set the top $k\%$ for one year and counted the number of individuals $n$ in that ranking. Then, for the previous year, we took the top-ranked individuals until we reached the same number $n$. Then we computed the number of individuals that entered or left the ranking as the flux values. Our results are for different percentiles: $1\%, 5\%,$ and $10\%$ in \sectionname~\textbf{S5.2}.

\subsection*{Matching strategies}

To examine the relationship between research productivity (number of papers) and citation impact (number of citations) among highly productive women and men researchers, we constructed three complementary samples designed to support robust comparisons. Researchers were grouped by field and career stage, defined as $2$--$10$, $11$--$25$, or $>25$ years since first publication. Highly productive researchers were identified as those in the top $k\%$ of productivity within each field and career stage, where $k$ equals 5 or 10. All estimates reported below are therefore conditional on having already reached the top $k\%$ of productivity. They describe citation differences \textit{among} researchers who occupy this group; they do not identify gender differences in the probability of reaching it, nor do they generalise to the broader population of researchers in each field.

\begin{figure}[!ht]
    \centering
   \includegraphics[width=0.8\linewidth]{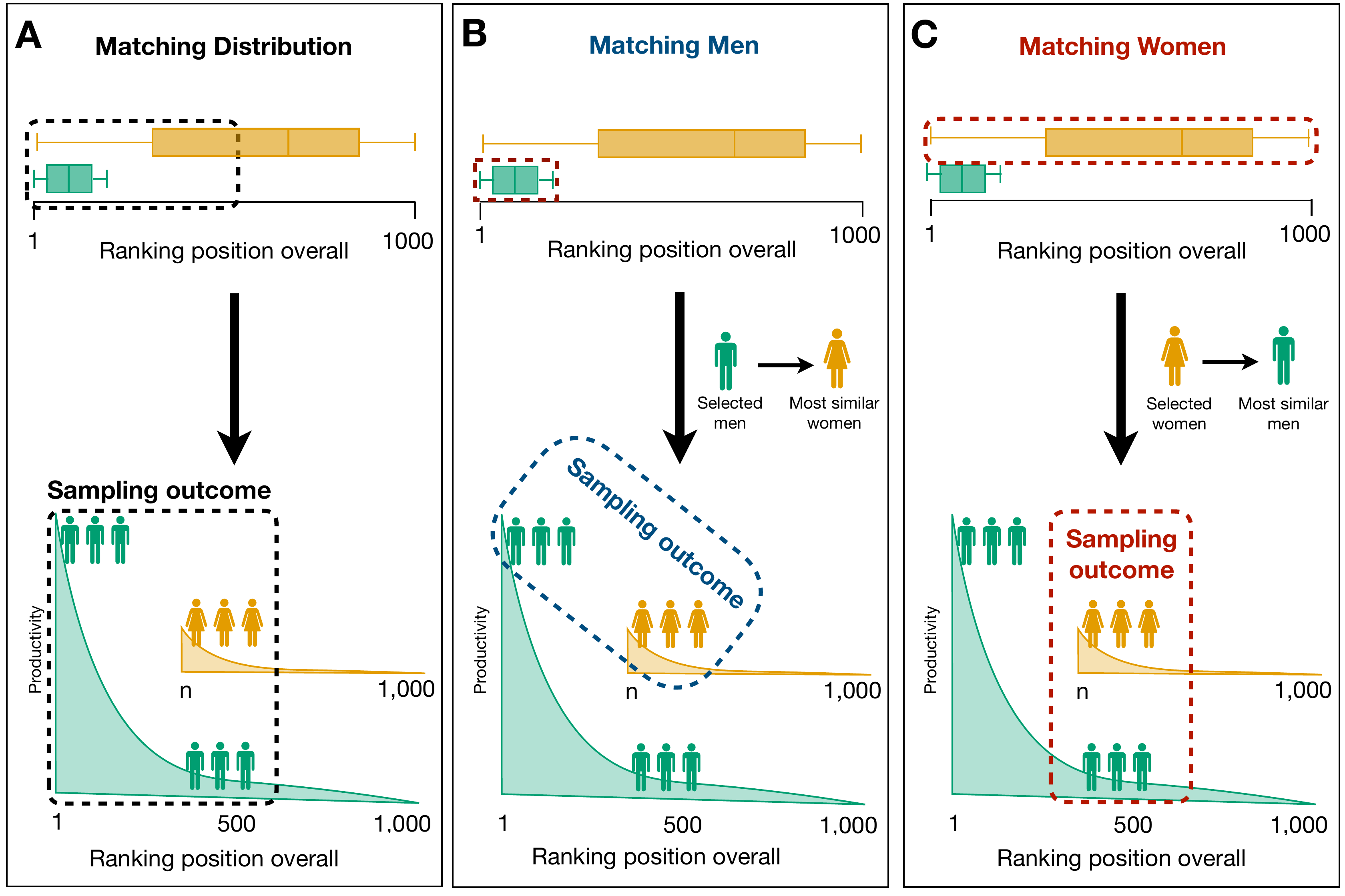}
    \caption{\textbf{Toy diagram of matching strategies.}  
    \textbf{A} Matching Distribution: top-productive (k\%) women and men are independently selected within each field and career age. \textbf{B} Matching Men: each top man is matched to the most similar woman, which often results in including top-ranked women even though this is not required. \textbf{C} Matching Women: each top woman is matched to the most similar man, who is not necessarily from the top ranks. Boxplots and distributions are only toy examples and do not correspond to any specific field.
    } 
    \label{fig:diagram_matching_strategies}
\end{figure}

For the first strategy, \matchingdistribution\ represented the observed composition of highly productive cohorts within each field and career stage, selecting researchers in the top $k\%$ of productivity whilst preserving the observed gender distribution. This strategy provided a baseline comparison among researchers who already occupied highly productive positions and corresponded to a common distribution-preserving benchmark in the fairness literature~\cite{zehlike2021fairness}. This strategy did not perform individual matching; instead, it retained the empirical structure of the observed top-productivity group, meaning that the numbers of women and men researchers differes. This strategy corresponds to the toy example shown in \figurename~\ref{fig:diagram_matching_strategies}\textbf{A}.

For the second and third strategies, we constructed individual matched comparisons. In these strategies, we anchored the sample on one gender group and identified the closest available counterpart of the other gender. Similarity was measured using a 13-dimensional publication profile comprising career age and twelve productivity counts defined by four field-specific venue citation quartiles and three authorship positions (first, middle, and last author). For example, these counts included the number of Q1 first-author papers, Q1 last-author papers, and Q2 last-author papers.

Authorship position was defined from the ordered author list: solo-authored papers were coded as first-author papers, and in two-author papers the second author was coded as last author. Venue quartiles were constructed from field- and year-specific venue PageRank computed from the venue directed citation network as a proxy for venue prestige, with Q1 corresponding to the highest-ranked venues and Q4 to the lowest-ranked venues (see \sectionname~\textbf{S6.1} for details). The productivity counts by venue quartile were constructed from papers with available venue information; weighted by the number of papers, venue information covered 98.2\% of the top 5\% samples and 97.5\% of the top 10\% samples across the three comparison strategies (see \sectionname~\textbf{S6.2} for field- and strategy-specific coverage).

Within each field and career stage, we computed Mahalanobis distances between researchers of different genders and selected the nearest available counterpart. This procedure identified researchers with the most similar observed publication profiles, without requiring them to occupy the same productivity ranking. Matches were one-to-one, so each researcher could be used only once; when multiple available counterparts shared the same minimum distance, one of the tied counterparts was selected at random. The main analysis includes all matched pairs: every anchored researcher was retained and paired with the nearest available counterpart, so the matched samples preserve, rather than discard, cases with limited overlap. The availability of comparable counterparts is part of what we aim to characterise when examining citation differences across the full reference group. At the same time, this implies that the degree of similarity achieved by the matching procedure may vary across pairs and strategies, making match quality important for interpreting the corresponding citation estimates. We therefore evaluated match quality using Mahalanobis distances, post-matching balance diagnostics on all 13 matching variables, and a distance-based sensitivity analysis that imposes progressively stricter calipers by excluding the most distant matched pairs, re-estimating both the models and the balance diagnostics within each restricted sample (see \sectionname~\textbf{S6.2}, \tablename~\textbf{S7}, and \sectionname~\textbf{S6.8}). For \matchingmen, we also conducted a distance-restricted sensitivity analysis, excluding the 0\%, 5\%, 10\%, 15\%, and 20\% of pairs with the highest Mahalanobis distances within each field and re-estimating both the models and balance diagnostics within each retained sample (Section~\textbf{S6.8}; Table~\textbf{S60}).

\begin{description}
    \item \matchingmen: This strategy anchors the comparison on men in the top \(k\%\) of productivity. Each highly productive man is matched to the woman with the smallest Mahalanobis distance in the same 13-dimensional publication profile, within the same field and career stage. This strategy identifies the closest available women counterparts to highly productive men and allows us to characterise how close the nearest available women counterparts are for this reference group. It corresponds to the toy example shown in~\figurename~\ref{fig:diagram_matching_strategies}\textbf{B} and also produces equal numbers of women and men.
    \item \matchingwomen: This strategy anchors the comparison on women in the top \(k\%\) of productivity. Each highly productive woman is matched to the man with the smallest Mahalanobis distance in the 13-dimensional publication profile, within the same field and career stage. This strategy identifies the closest available men counterparts to highly productive women, even when those men do not necessarily occupy similar ranking positions. It corresponds to the toy example shown in~\figurename~\ref{fig:diagram_matching_strategies}\textbf{C} and produces equal numbers of women and men.
\end{description}

These three matching strategies capture complementary comparative scenarios. \textit{i)} The \matchingdistribution\ strategy reflects the actual distribution of performance across each field and career stage, preserving the observed gender proportions. \textit{ii)} The \matchingmen\ strategy anchors on the top-performing men and matches each to the most similar women, making explicit the distance between women’s observable profiles and the benchmarks defined by the highest-ranked men. \textit{iii)} The \matchingwomen\ strategy anchors on the top-performing women and matches each to the most similar men, not necessarily selecting men at the highest ranks.

If men and women were symmetric in their observable distributions, all three strategies would converge to similar matched samples and comparable results. Their divergence instead reveals different dimensions of structural asymmetry: \matchingdistribution\ provides the baseline disparities observed in real cohorts, \matchingmen\ emphasises the differences that remain when men’s ranking advantage is explicitly retained, and \matchingwomen\ shows how gaps change when women are compared to equivalent men without considering their ranking. Taken together, these perspectives provide a systematic framework to examine gender disparities at the top of academic recognition. For more details on the matching quality using the Mahalanobis distances across fields and strategies, see \sectionname~\textbf{S6.6}. 

\subsection*{Negative Binomial Fixed Effects Regression Modelling}
\subsubsection*{Model choice and diagnostics}
\label{sec:methods_reg}

The Negative Binomial (NB) Linear Model is a statistical method designed to analyse count data that exhibit overdispersion. This approach is particularly suitable for our study, as it models discrete outcomes such as papers, citations, and co-authors, and is preferred over Poisson models when the variance substantially exceeds the mean—a common feature of citation data given their heavy-tailed distributions (Figure~\ref{fig:Representation_top_ranked}). 

To verify that the NB model is appropriate in our context, we conduct several diagnostic tests (\sectionname~\textbf{S6.4}). The Cameron–Trivedi test consistently rejects the null hypothesis of equidispersion ($p<0.05$ in most cells), indicating that variance exceeds the mean and that Poisson models are misspecified. The estimated dispersion parameter (theta) was finite and clearly greater than zero across specifications, further confirming the presence of overdispersion. Finally, model fit comparisons based on AIC showed that NB regressions consistently outperform Poisson alternatives.

Together, these diagnostics strongly support using NB regression as the primary modelling approach. For completeness, we also estimated Poisson models, and the results are qualitatively consistent with those from the NB regressions (\tableautorefname~\textbf{S19}–\textbf{S20}).

\subsubsection*{Model specification}  

For each matching strategy and productivity top-$k$ threshold, we estimated Negative Binomial fixed-effects models on an unbalanced panel of researcher--year observations. The dependent variable is the cumulative number of citations researcher $i$ received in year $t$. The main specification is:

\begin{equation}
\begin{aligned}
\mathit{Citations}_{it} &\sim \operatorname{NegBin}(\mu_{it}, \theta) \\[2mm]
\log(\mu_{it}) &= \beta_0
+ \beta_1\,\mathit{Gender}_{i}
+ \sum_{d \in D} \beta_{2d}\,\mathit{CareerStartDecade}_{id}
+ \beta_3\,\mathit{CareerMedium}_{it}
+ \beta_4\,\mathit{CareerSenior}_{it} \\[1mm]
&\quad + \sum_{q \in Q}\sum_{p \in P} \beta_{5qp}\,\mathit{Productivity}_{itqp}
+ \beta_6\,\mathit{Coauthorships}_{it} \\[1mm]
&\quad + \beta_7\,\bigl(\mathit{TotalProductivity}_{it} \times \mathit{Coauthorships}_{it}\bigr)
+ \tau_t \;[+\;\delta_{f(i)} \text{ in pooled models}].
\end{aligned}
\label{eq:main_nb_model}
\end{equation}
$\mathit{TotalProductivity}_{it}$ is defined as the sum of the twelve productivity cells included in the model:
\begin{equation*}
\mathit{TotalProductivity}_{it} = \sum_{q \in Q}\sum_{p \in P} \mathit{Productivity}_{itqp},
\end{equation*}
where $Q=\{\mathrm{Q1},\mathrm{Q2},\mathrm{Q3},\mathrm{Q4}\}$ denotes venue quartiles and $P=\{\text{first},\text{middle},\text{last}\}$ denotes authorship positions.

Venue quartiles were constructed within each field and year by computing PageRank on directed venue-to-venue citation networks and splitting venues into Q1--Q4 categories; each paper then inherited the quartile of its publication venue in that year. Thus, productivity in the regression was not treated as a single publication count, but as twelve cumulative counts combining venue quartile and authorship position.

In \equationautorefname~\ref{eq:main_nb_model}, $\mu_{it}$ is the expected cumulative citations of researcher $i$ at time $t$, and $\theta$ is the NB overdispersion parameter. The term $\mathit{Gender}_i$ is a binary indicator equal to one for women and zero for men. The $\mathit{CareerStartDecade}_{id}$ terms are dummy variables for the decade in which researcher $i$ started publishing, with the pre-1980 cohort omitted as the reference category; $D$ denotes the set of included career-start decade categories. Career stage was included as a time-varying indicator based on years since first publication, following prior literature~\cite{jadidi2018gender,jaramillo2021reaching} and detailed in \sectionname~\textbf{S3}: Early ($2$--$10$ years), Medium ($11$--$25$ years), and Senior ($> 25$ years), with Early as the reference category. The term $\mathit{Productivity}_{itqp}$ is the number of publications by researcher $i$ up to year $t$ in venue quartile $q$ and authorship position $p$. Finally, $\tau_t$ denotes year fixed effects, accounting for time-specific factors affecting all researchers in a given year; pooled models additionally included field fixed effects, $\delta_{f(i)}$.

We estimated two families of models. \textit{i)} \textit{Field-specific regressions}, estimated separately for each discipline, allowed coefficients to be interpreted relative to disciplinary norms and practices. \textit{ii)} \textit{Pooled regressions}, estimated across all fields with field fixed effects, captured overarching patterns that persisted across fields. In the results, this pooled specification corresponds to the \texttt{All fields} category in Figures~\ref{fig:gender_disparities_regressions_5pct} and~\textbf{S40}.

For inference, we report Driscoll--Kraay standard errors with researcher as the panel unit and year as the time index, because the diagnostics reported in the Supplementary Material indicate heteroskedasticity, serial correlation, and cross-sectional dependence.

In \equationautorefname~\textbf{S1} we show an expanded version of \equationautorefname~\ref{eq:main_nb_model}, writing the twelve productivity terms separately by venue quartile and authorship position. As an additional robustness check, we also re-estimated the models using cumulative citations excluding self-citations as an alternative outcome with qualitatively similar results (see \sectionname~\textbf{S6.6}).

\subsubsection*{Interpretation of coefficients}

For clarity, regression coefficients are reported as incidence rate ratios (IRR) and expressed as percentage changes in expected citations, $100\times[\exp(\beta)-1]$. This presentation is applied consistently across the main text (\eg \figurename~\ref{fig:gender_disparities_regressions_5pct}) and the Supplementary Material (\eg \figurename~\textbf{S40}, and Tables~\textbf{S22}--\textbf{S59}).

Specifically, for the productivity terms and coauthorships, effects are expressed as $(\exp(\beta_{5qp}) - 1) \times 100\%$ and $(\exp(\beta_6) - 1) \times 100\%$, respectively. For example, the productivity effect shown in Figure~\ref{fig:gender_disparities_regressions_5pct} corresponds to the transformed coefficient for Q1 first-author productivity. For \textit{Gender}, because $\mathit{Gender}_i$ is coded as one for women and zero for men, $\exp(\beta_1)$ compares women with men as the reference group. We therefore express the gender coefficient as $(\exp(\beta_1)-1)\times100\%$: positive values indicate higher expected citations for women relative to men, whereas negative values indicate lower expected citations for women relative to men.


\newpage
\bibliography{sn-bibliography}



\section*{Supplementary information}

All supplementary information is detailed in an external PDF.

\section*{Declaration}

For the purpose of open access, the authors have applied a Creative Commons Attribution (CC BY) license to any accepted manuscript version arising from this submission. 

\section*{Ethics \& Inclusion statement}
We declare that principles of ethics \& inclusion were considered during the design, development, and analysis of this research. Our results do not intend to be discriminatory or stigmatising to researchers of any gender or sex, but rather highlight the unequal participation and representation of certain groups of researchers and the possible related causes. Corresponding limitations of the gender inference methods used in the analyses are disclosed in the discussion of the manuscript.

\section*{Data availability statement}
The datasets analysed during the current study are available in the Semantic Scholar Open Research Corpus\\
\href{https://github.com/allenai/s2orc}{\textcolor{blue}{https://github.com/allenai/s2orc}}. Gender inference was performed using Genderize~\cite{genderize} and Namsor~\cite{Namsor}, but this information is not publicly available to maintain the privacy of individuals' gender disclosure. 

\section*{Code availability statement}
The code used in the current study does not involve the development of new software or algorithms, nor the proof of mathematical theories through computational modelling; instead, it utilises statistical methods or metrics previously defined in the literature, as stated in the manuscript. The code to generate the plots displayed in the main manuscript can be found in this \href{https://github.com/anajaramillo37/Systematic-comparison-of-gender-inequality-in-scientific-rankings-across-disciplines}{GitHub repository}.

\section*{Acknowledgments}

We extend our gratitude to Semantic Scholar for providing access to a comprehensive dataset of publications across multiple fields. We thank the Max Planck Institute for Demographic Research (MPIDR) for their constructive feedback during the Scholarly Migration and Mobility Symposium. Our gratitude also extends to Dr. Julia Sachseder, who read early versions of the manuscript and provided us with literature from the fields of gender studies and organisational sciences.

Ana Maria Jaramillo and Fariba Karimi acknowledge the support of the MAMMOTH project Horizon 2020 HORIZON-CL4-2021-HUMAN-01-24 (RIA). Mariana Macedo acknowledges the partial financial support of the Center for Collective Learning led by Dr. César A. Hidalgo and the Artificial and Natural Intelligence Institute of the University of Toulouse ANR-19-PI3A-0004. Melanie Oyarzun was funded by the European Union under the Horizon EU project LearnData (101086712).

\section*{Author contributions statement}

All authors (AJ, MM, MO, MAO, FK and RM) contributed to the original idea; AJ and MM designed the study and performed the analysis; RM and FK supervised the development of the experiments; AJ, MM and MO prepared the analyses, tables and graphics; MO and MM conducted econometric analyses to support the findings; All authors read, wrote, reviewed, and approved the final manuscript.




\end{document}